\documentclass[aps,prb,superscriptaddress,floatfix]{revtex4}
\usepackage{epsfig,graphicx,times}
\begin{document}

\title{Errata: Diffusion Monte Carlo study of circular quantum dots\\
$[$Phys. Rev. B 62, 8120 (2000)$]$}

%\author{Francesco Pederiva}
%\address{Dipartimento di Fisica and INFM, Universit\`a di Trento, I-38050 Povo,Trento, Italy}
%\author{C. J. Umrigar}
%\address{Cornell Theory Center, Cornell University, Ithaca, NY 14853}
%\author{E. Lipparini}
%\address{Dipartimento di Fisica and INFM, Universit\`a di Trento, I-38050 Povo,Trento, Italy}

\author{Francesco Pederiva, C. J. Umrigar and E. Lipparini}

\date{\today}
%\pacs{73.21.La}
\maketitle

Several of the energies in Tables I and II are incorrect.  Most of the errors are due to
incorrectly inputting the symmetry of some of the states, others are due to incorrectly
converting or transcribing the energies.  In particular, for $N=4$ we had
a near degeneracy and a violation of Hund's first rule.  The corrected result has $|L=0,S=1\rangle$
as the ground state, as predicted by Hund's rule.  Hund's first rule is satisfied for all $N$
for the confining potential used, but there is a near degeneracy for $N=10$.
We present corrected versions of Tables I and II.  All quantum Monte Carlo results have
been recomputed with an improved Jastrow factor.
All LSDA values in Table II have also been recalculated.
The rms fluctuations of the local energy in VMC
range from 0.008 H$^*$ for $N=2$ to 0.24 H$^*$ for the $N=13$, $|L=1,S=1/2\rangle$ state.
The number of determinants, $N_{\rm det}$, in the wave functions depends on whether real or complex
orbitals are employed.  The values shown in Table I are for real orbitals.
Not only the DMC energies, but also the VMC energies computed with LSDA and LDA orbitals
agree within 1  mH$^*$ in all cases tested.
All LSDA values in Table II have also been recalculated.
The effective Bohr radius $a_0^*$ should be 97.9373 \AA, rather than 97.93 \AA.

\begin{table}
\caption{Ground state energies (in H$^*$) and
low-lying excitation energies (in mH$^*$) for $N\le13$ dots.
Also shown are the quantum numbers of the states and
the number of configuration state functions $N_{\rm conf}$ and the
number of determinants $N_{\rm det}$ used in constructing them.
The numbers in parentheses are the statistical uncertainties in the last digit.
}
\begin{center}
\begin{tabular}{cccccr}
 \toprule
$N$ & $L$ & $S$ & $N_{\rm conf}$ & $N_{\rm det}$ & $E$(H$^*$), $\Delta E$(mH$^*$)\\
 \colrule
 2 & 0 & 0   & 1 & 1 & 1.02164(1)\\
 \colrule
 3 & 1 & 1/2 & 1 & 1 & 2.2339(1)\\
 \colrule
 4 & 0 & 1   & 1 & 1 & 3.7145(1)\\
   & 2 & 0   & 1 & 2 & 41.1(1)\\
   & 0 & 0   & 1 & 2 & 66.3(1)\\
 \colrule
 5 & 1 & 1/2 & 1 & 1 & 5.5338(1)\\
 \colrule
 6 & 0 & 0   & 1 & 1 & 7.6001(1)\\
 \colrule
 7 & 2 & 1/2 & 1 & 1 & 10.0342(1)\\
   & 0 & 1/2 & 1 & 1 & 27.5(1)\\
 \colrule
 8 & 0 & 1   & 1 & 1 & 12.6900(1)\\
   & 2 & 1   & 1 & 1 & 21.9(1)\\
   & 4 & 0   & 1 & 2 & 27.5(1)\\
   & 0 & 0   & 2 & 3 & 36.0(1)\\
   & 2 & 0   & 1 & 2 & 56.1(1)\\
 \colrule
 9 & 0 & 3/2 & 1 & 1 & 15.5801(1)\\
   & 2 & 1/2 & 2 & 2 & 28.5(1)\\
   & 4 & 1/2 & 1 & 2 & 42.6(1)\\
   & 0 & 1/2 & 2 & 5 & 55.1(1)\\
 \colrule
10 & 2 & 1   & 1 & 1 & 18.7232(1)\\
   & 0 & 0   & 2 & 3 &  2.9(1)\\
   & 0 & 1   & 1 & 1 & 23.3(1)\\
   & 2 & 0   & 1 & 2 & 40.0(1)\\
   & 4 & 0   & 1 & 2 & 46.7(1)\\
 \colrule
11 & 0 & 1/2 & 1 & 1 & 22.0738(1)\\
   & 2 & 1/2 & 1 & 1 & 15.3(1)\\
 \colrule
12 & 0 & 0   & 1 & 1 & 25.6356(1)\\
 \colrule
13 & 3 & 1/2 & 1 & 1 & 29.4938(1)\\
   & 1 & 1/2 & 1 & 1 & 39.2(1)\\
 \botrule
\end{tabular}
\end{center}
\label{tab.excited}
\end{table}

\begin{table}
\caption[]{Comparison of ground state energies (in H$^*$) for the dots with $2\leq N\leq13$
computed by Hartree--Fock, LSDA, VMC and DMC. Also shown are the LSDA errors
in the energy, $\Delta E_{\rm LSDA} = E_{\rm LSDA}-E_{\rm DMC}$, which
are much smaller than the HF errors $E_{\rm HF}-E_{\rm DMC}$.
The numbers in parentheses are the statistical uncertainties in the last digit.
}
\begin{center}
\begin{tabular}{rrrrrr}
\toprule
N&$E_{\rm HF}$&$E_{\rm LSDA}$&$E_{\rm VMC}$&$E_{\rm DMC}$&$\Delta E_{\rm LSDA}$\\
\colrule
 2& 1.1420& 1.04684& 1.02165(1)& 1.02164(1)&0.02520(1)\\
 3& 2.4048& 2.26308& 2.2395(1)& 2.2339(1)&0.0292(1)\\
 4& 3.9033& 3.74632& 3.7194(1)& 3.7145(1)&0.0318(1)\\
 5& 5.8700& 5.56919& 5.5448(1)& 5.5338(1)&0.0354(1)\\
 6& 8.0359& 7.63500& 7.6104(1)& 7.6001(1)&0.0349(1)\\
 7&10.5085&10.07176&10.0499(1)&10.0342(1)&0.0376(1)\\
 8&13.1887&12.72691&12.7087(1)&12.6900(1)&0.0369(1)\\
 9&16.1544&15.61889&15.5996(1)&15.5801(1)&0.0388(1)\\
10&19.4243&18.76357&18.7496(1)&18.7232(1)&0.0404(1)\\
11&22.8733&22.11130&22.1018(1)&22.0738(1)&0.0375(1)\\
12&26.5490&25.67597&25.6659(1)&25.6356(1)&0.0404(1)\\
13&30.4648&29.53617&29.5295(1)&29.4938(1)&0.0424(1)\\
\botrule
\end{tabular}
\end{center}
\label{tab.ground}
\end{table}

\end{document}